\documentclass[aps,prd,twocolumn,showpacs,superscriptaddress,amsmath,graphicx,psfrag,color,longtable]{revtex4}

\usepackage{graphicx}
\usepackage{psfrag}
\usepackage{epsfig}
\usepackage{color}
\usepackage{longtable}
\usepackage{amsmath}
\usepackage{amssymb}

\newcommand{\dd}{\mathrm{d}}
\newcommand{\sy}{\hat{s}}
\newcommand{\uy}{\hat{u}}

\newcommand{\ii}{\mathrm{i}}

\newcommand{\Leff}{{\cal L}_{\mathrm{eff}}}

\def\slashed#1{\displaystyle{\not}#1}

\definecolor{Red}{rgb}{1.,0.,0.}

\begin{document}

\title{Signatures of NP models in top FCNC decay $t\to c(u) \ell^+\ell^-$}

\author{Jure Drobnak}
\email[Electronic address:]{jure.drobnak@ijs.si}
\affiliation{J. Stefan Institute, Jamova 39, P. O. Box 3000, 1001 Ljubljana, Slovenia}

\author{Svjetlana Fajfer}
\email[Electronic address:]{svjetlana.fajfer@ijs.si}
\affiliation{J. Stefan Institute, Jamova 39, P. O. Box 3000, 1001 Ljubljana, Slovenia}
\affiliation{Department of Physics, University of Ljubljana, Jadranska 19, 1000 Ljubljana, Slovenia}

\author{Jernej F. Kamenik}
\email[Electronic address:]{jernej.kamenik@ijs.si}
\affiliation{J. Stefan Institute, Jamova 39, P. O. Box 3000, 1001 Ljubljana, Slovenia}
\affiliation{INFN, Laboratori Nazionali di Frascati, I-00044 Frascati, Italy}

\date{\today}

\begin{abstract}
Among other roles the LHC will play the role of a ``top factory'' giving us an unique possibility to study possible new physics signatures in unprecedented ways.
Many scenarios of new physics (NP) allow top quark flavour changing neutral current (FCNC) decays. Using the most general model independent Lagrangian we investigate possible experimental signals
of new physics in $t\rightarrow c(u)l^+l^-$ FCNC top decays. We find that a measurement of two possible angular asymmetries might give very important and interesting information on the structure of NP contributions. It is particularly interesting to use these observables to discriminate among variety of NP scenarios. Among others, we consider contributions due to the interference between scalar and vector mediators of these FCNC decays.
\end{abstract}


\maketitle

\section{Introduction}

Top quark physics plays an important role in the present era of Tevatron and LHC experiments. The high mass of the top quark offers a much richer phenomenology compared to other, lighter SM fermions. In order to stabilize the Higgs mass new physics (NP) is expected to arise and possibly to induce new flavour structures. It is rather well known that the standard model (SM) predicts highly suppressed effects of flavour changing neutral currents (FCNC) and that physics beyond the SM (BSM) in many cases lifts this suppression (for a recent review c.f.~\cite{AguilarSaavedra:2004wm}). Top FCNCs can be searched for both in production and decays (for current Tevatron limits c.f.~\cite{Abazov:2007ev, :2008aaa}). The LHC can be considered as a ``top factory'', producing about 80,000 $t\bar t$ events per day at the luminosity $L=10^{33}~\mathrm{cm}^{-2}\mathrm{s}^{-1}$ and being able to access rare top decay branching ratios at the $10^{-5}$ level~\cite{Jana:2008kb}.

FCNC top decays can be approached within some of many specific models (c.f. refs.~[3] of~\cite{Fox:2007in}). Another possibility is to use a model independent analysis (c.f. refs.~[4] of~\cite{Fox:2007in}) which can then also be applied to concrete model implementations. In the present work, we apply this strategy. Usually, theoretical studies make predictions for the branching ratios
$t \to c (u) Z(\gamma,h)$, for which there already exist feasibility and sensitivity studies for the LHC experiments~\cite{Carvalho:2007yi, VanderDonckt:2008ax}. Our analysis on the other hand is devoted to the study of $t \to c (u) l^+ l^-$ with the basic goal of identifying discriminating
effects of different NP models in top FCNCs which can be approached by the
experimental study. Exploring the three-body decay channel offers three advantages: (1) the larger phase-space offers more observables to be considered -- in particular the \emph{angular asymmetries among the final lepton and jet directions}; (2) the channel may receive contributions from BSM particle mediation, such as new heavy scalar or vector resonances; (3) many models predict observable effects in more than one top FCNC two-body decay mode making interference effects in the common three-body channel important.

Since the standard forward-backward asymmetry for the leptons vanishes in the photon and scalar mediated decays, we consider another asymmetry which we call the left-right asymmetry and is associated with the lepton angular distribution in the lepton-quark rest frame (see section \ref{asim}). This asymmetry is nontrivial also in the photon and scalar mediated decays.
We explore the ranges of values for these two asymmetries in $t\rightarrow c(u)l^+l^-$ decays mediated by different bosons. We also consider certain interferences between them as we expect these to significantly affect the asymmetries.
Our results can serve as a starting point for more elaborate investigations of experimental sensitivity to the proposed observables including QCD corrections, proper jet fragmentation and showering and the impact of experimental cuts and detector effects.

The paper is structured as follows: In section II we introduce the effective top FCNC Lagrangian and set our notation, while in section III we discuss the possible observables in $t \to c (u) l^+ l^-$ decay. Section IV contains a model independent analysis of possible distinguishable NP signatures using these observables. Conclusions with outlook are summarized in section V. 

\section{Effective Lagrangian}

We consider the most general effective Lagrangian describing $t\to c\ell^+\ell^-$ transitions mediated by SM gauge fields ($\gamma, Z$)\footnote{We do not consider FCNC top-gluon interactions, as they do not contribute to our final state at tree-level. They would however need to be included in a calculation of $\alpha_s$ corrections.}, a light (SM-like) Higgs field ($\phi$) as well as possible contributions due to exchange of heavy scalar or vector resonances ($\phi', Z'$) in both $s$- and $u$-channels. 

\subsection{FCNC mediation by SM fields}

In writing the effective top FCNC Lagrangian we follow roughly the notation of ref.~\cite{AguilarSaavedra:2004wm, AguilarSaavedra:2008zc}. 
Hermitian conjugate operators are assumed implicitly to be contained in the Lagrangian and contributing to the charge conjugated decay modes
\begin{eqnarray}
\mathcal L^{tc}_{\mathrm {eff}} &=& \mathcal L^{Z}_{\mathrm {eff}} + \mathcal L^{\gamma}_{\mathrm {eff}} + \mathcal L^{\phi}_{\mathrm {eff}}\,,\nonumber\\
\mathcal L^{Z}_{\mathrm {eff}} &=& g_Z \frac{v^2}{\Lambda^2} Z_{\mu} \left[ a_L^Z \bar q_L \gamma^{\mu} t_L + a_R^Z \bar q_R \gamma^{\mu} t_R\right]\nonumber\\
&& +  g_Z \frac{v}{\Lambda^2} Z_{\mu\nu} \left[ b_{RL}^Z \bar q_R \sigma^{\mu\nu} t_L + b_{LR}^Z \bar q_L \sigma^{\mu\nu} t_R\right]\,,\nonumber\\
\mathcal L^{\gamma}_{\mathrm {eff}} &=& e \frac{v}{\Lambda^2} A_{\mu\nu} \left[ b_{RL}^\gamma \bar q_R \sigma^{\mu\nu} t_L + b_{LR}^\gamma \bar q_L \sigma^{\mu\nu} t_R\right]\,,\nonumber\\
\mathcal L^{\phi}_{\mathrm {eff}} &=& \frac{v^2}{\Lambda^2} \phi \left[ c_{RL}^h \bar q_R t_L + c_{LR}^h \bar q_L t_R\right]\,,
\label{eq:L-SM}
\end{eqnarray}
where $q=c(u)$, $q_{R,L} = (1\pm\gamma_5)q/2$, $\sigma_{\mu\nu} = i[\gamma_\mu,\gamma_\nu]/2$, $g_Z = 2 e/\sin 2 \theta_W$ and $X(A,Z)_{\mu\nu} = \partial_{\mu} X_\nu - \partial_\nu X_\mu$, 
$v$ is the vacuum expectation value (VEV) of the SM scalar $SU(2)$ doublet and $\Lambda$ is the scale of new physics.

We assume strictly SM-like lepton flavour conserving interactions, as these have been precisely measured (with the exception of the Higgs couplings) by low energy experiments and at LEP
\begin{eqnarray}
\mathcal L^{\ell}_{\mathrm {eff}} &=& g_Z Z_{\mu} \left[ \sin^2\theta_W \bar \ell_R \gamma^{\mu} \ell_R - (\cos 2\theta_W /2) \bar \ell_L \gamma^{\mu} \ell_L\right]\nonumber\\
&& + e A_{\mu} \bar \ell \gamma^{\mu} \ell + (m_\ell / v) \phi \left[ \bar \ell_R \ell_L + \bar \ell_L \ell_R\right]\,.\label{eq:L-SM2}
\end{eqnarray}

\subsection{FCNC mediation by heavy resonances}

In addition to $t\to c(u) \ell^+\ell^-$ mediation by SM fields, we also consider possible contributions due to the exchange of new BSM heavy scalar ($\phi'$) or vector ($Z'$) resonances. FCNC couplings in the up-quark sector can lead to contributions in the $s$-channel
\begin{eqnarray}
\mathcal L^{tc}_{\mathrm {eff}} & \rightarrow &\mathcal L^{tc}_{\mathrm {eff}}+ \mathcal L^{Z'}_{\mathrm {eff}} + \mathcal L^{\phi'}_{\mathrm {eff}}\,,
\end{eqnarray}
where the explicit expressions can be obtained from eq.~(\ref{eq:L-SM}) via substitution $Z\to Z'$ and $\phi\to \phi'$ in the two new terms.
In this case we only need to consider effective scalar and vector current flavour conserving interactions for the lepton sector
\begin{eqnarray}
\mathcal L^{\ell}_{\mathrm {eff}} &\rightarrow &\mathcal L^{\ell}_{\mathrm {eff}}+ g_Z Z'_{\mu} \left[ a^{\ell}_R \bar \ell_R \gamma^{\mu} \ell_R +a_L^{\ell} \bar \ell_L \gamma^{\mu} \ell_L\right]\nonumber\\
&& + \phi' \left[ c_{RL}^{\ell} \bar \ell_R \ell_L + c_{LR}^{\ell}  \bar \ell_L \ell_R\right]\,.
\end{eqnarray}

On the other hand, baryon and lepton number violating interactions of charged heavy resonances may lead to contributions via the $u$-channel exchange. In the spin-averaged decay observables we consider, this case can be fully taken into account by interchanging the four-momentum labels of the light-quark jet and the positive signed lepton. In term, this leads to interchange of angular and charge asymmetries, which we consider in the next section.

Finally we would like to note that in the limit $m_{\phi',Z'}\to \infty$ while keeping the product $m_{\phi',Z'} \Lambda$ constant and finite we can also address the possibility of effective $\bar c t \ell^+ \ell^-$ contact interactions due to non-perturbative or loop induced NP contributions.

\section{Observables}

We consider scenarios where detection of a NP signal in the FCNC decay channel $t\to c (u) \ell^+\ell^-$ could be most easily complemented by other observables in the same decay mode. This would allow distinguishing between different possible effective amplitude contributions and thus different underlying NP models. Therefore we consider the most inclusive (fully phase-space integrated and spin-averaged) observables, complementary to the branching ratio (analytic formulae for the relevant partly-integrated decay distributions are given in the appendix). In the present work, we give formulae, calculated at the parton level and with minimal experimental cuts imposed on the kinematic variables. Finally, we neglect kinematical effects of lepton masses and the light quark jet invariant mass, as these are expected to yield immeasurably small effects in the kinematical phase-space set by the large top quark mass. 

\subsection{Differential decay rates and branching ratio}

We start with the double-differential decay rate ${\dd \Gamma}/{(\dd u \dd s)}$, where $s=m_{l^+l^-}^2$ is the invariant mass of the lepton pair and
$u = m_{j l^+}^2$ is the invariant mass of the final state quark (jet) and the lepton of positive charge $l^+$. Integrating this decay rate over one of the kinematical variables, we obtain
the partially integrated decay rate distributions ($\dd\Gamma / \dd u$, $\dd\Gamma / \dd s$), while the full decay rate ($\Gamma$) is obtained after completely integrating these distributions. The branching ratio is usually obtained by normalizing to the dominating charged current $t\to b W$ decay width.

\subsection{Angular or charge asymmetries}
\label{asim}

\begin{figure}[h]
\begin{center}
\includegraphics[scale=0.7]{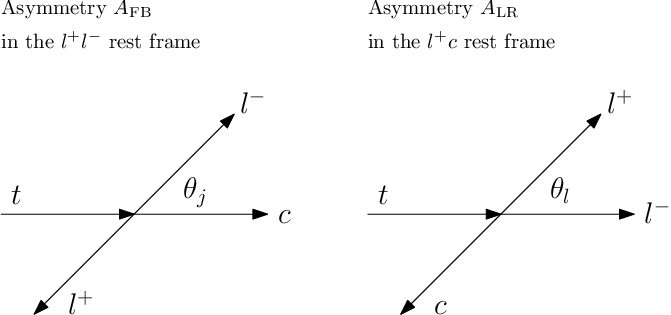}
\caption{\label{fig:angles}\small Definition of two angles relevant to our analysis. }
\end{center}
\end{figure}

The differential decay rate distribution can also be decomposed in terms of two independent angles, as defined in fig.~\ref{fig:angles}. 
In the $\ell^+ \ell^-$ rest-frame $z_j=\cos\theta_{j}$ measures the relative direction between the negatively charged lepton and the light quark jet. Conversely, in the rest-frame of the positive lepton and the quark jet, we can define $z_{\ell}=\cos\theta_{\ell}$ to measure the relative directions between the two leptons. In terms of these variables, we can define two asymmetries ($i=j,\ell$) as
\begin{equation}
A_i = \frac{\Gamma_{z_i>0}-\Gamma_{z_i<0}}{\Gamma_{z_i>0} + \Gamma_{z_i<0}}\,,
\label{eq:As}
\end{equation}
where we have denoted $\Gamma_{z_i\lessgtr 0}$ as the integrated decay rates with an upper or lower cut on one of the $z_i$ variables. We can then identify $A_{j}\equiv A_{\mathrm{FB}}$ as the commonly known forward-backward asymmetry (FBA) and define $A_{l}\equiv A_{\mathrm{LR}}$ as the left-right asymmetry (LRA). The two angles and the asymmetries they define are related via a simple permutation of final state momentum labels between the quark jet and the positively charged lepton, and consequently via a $u\leftrightarrow s$ interchange.
Finally, since the asymmetries as defined in eq.~(\ref{eq:As}) are normalized to the decay rate, they represent independent observables with no spurious correlations to the branching ratio. On the other hand, correlations among the two asymmetries are of course present and indicative of the particular NP operator structures contributing to the decay.

\section{Signatures}

Next we study the signatures of various possible contributions to the $t\to c(u) \ell^+\ell^-$ decay using the integrated observables defined in the previous section. 
Before exploring individual mediation cases a general remark is in order. Since all the effective dimension-five and -six operators in eq.~(1) come suppressed with an undetermined NP cut-off scale, the actual values of the effective couplings ($a_i$, $b_i$, $c_i$) are unphysical (can always be shifted with a different choice of the cutoff scale). The total decay rate determines the overall magnitude of the physical product of the couplings with the cut-off scale. On the other hand relative sizes or ratios of couplings (independent of the cut-off) determine the magnitude of the asymmetries. The extremal cases are 
then naturally represented when certain (combinations of) couplings are set to zero -- often the case in concrete NP model implementations.

\subsection{Photon mediation}

Usually, direct detection of energetic photons is considered to be the prime strategy in the search for photon mediated FCNCs of the top. However the $t\to c(u)\ell^+\ell^-$ channel can serve as an additional handle.
Due to the infrared pole in the di-lepton invariant mass distribution we introduce a low $\sy=m_{\ell}^2/m_t^2$ cut denoted $\sy_{\mathrm{min}}\equiv\epsilon/m_t^2$ and present the
total decay width as its function. The physical cut is of course at $\epsilon=4m_{\ell}^2$. We also define an auxiliary variable summarizing the relevant NP parameter dependencies
$$
B_{\gamma} = \frac{m_t^2}{v^2}\frac{e^4}{g_Z^4}\frac{|b_{LR}^{\gamma}|^2+|b_{RL}^{\gamma}|^2}{2}\,,
$$
in terms of which the fully integrated decay width is
\begin{eqnarray}
\Gamma &=& \frac{m_t}{16\pi^3}\frac{g_Z^4 v^4 }{\Lambda^4}B_{\gamma}f_{\gamma}(\epsilon)\,.
\end{eqnarray}
Function $f_{\gamma}$ depends only on the di-lepton invariant mass cutoff $\epsilon$ and is presented in eq.~(\ref{fg}) of appendix A.
The FBA vanishes identically, while for the LRA we obtain
\begin{eqnarray}
A_{\mathrm{LR}}&=&\frac{g_{\gamma}(\epsilon)}{f_{\gamma}(\epsilon)}\,.
\end{eqnarray}
The asymmetry does not depend on the effective dipole couplings, however there is a non-trivial dependence of the LRA on the low $\sy$ cut, which we plot in fig.~\ref{fig:lra_photon}.
The function $g_{\gamma}$ is presented in eq.~(\ref{gg}) of appendix A.
\begin{figure}[h] 
\begin{center}
\includegraphics[scale=1]{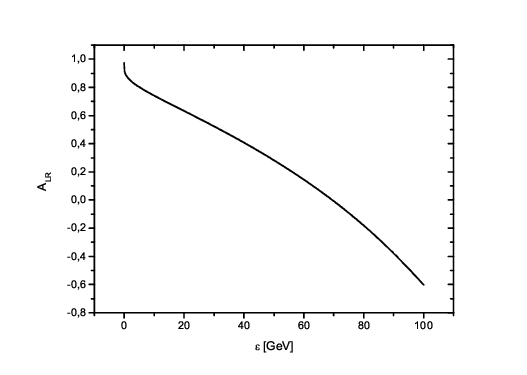}
\caption{\label{fig:lra_photon}\small The dependence of the photon-mediated LRA on the low di-lepton invariant mass cut $\epsilon$. }
\end{center}
\end{figure}
We see that the value as well as the sign of the integrated LRA is highly sensitive to the cut.

\subsection{$Z$ mediation}

\begin{figure}[h]
\begin{center}
\includegraphics[scale=1]{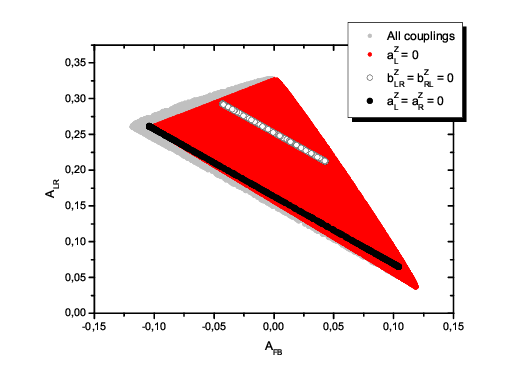}
\caption{\label{fig:assym_Z}\small The correlation of FBA and LRA in $Z$ mediated decay. The gray area represents decays with all possible current and dipole $Z$ FCNC couplings. The
red area corresponds to decays with $a_L^Z$ set to zero, while the white and black lines represent decays with only current and only dipole couplings respectively.}
\end{center}
\end{figure}

Current search strategies for $t\to c Z$ decays actually consider $t\to c \ell^+\ell^-$, but in addition impose a cut on the invariant lepton mass around the $Z$ mass to reduce backgrounds. As long as such cuts are loose compared to the width of the $Z$, we do not expect them to affect our observables. This is of course valid only if $Z$ is the only dominating FCNC mediation channel. Regarding the model dependent parameters, we denote $L_+$ and $L_-$ originating from
the $\mathcal L_{\mathrm{eff}}^l$ couplings
\begin{equation}
L_{\pm} = \frac{\sin^4\theta_W\pm\frac{1}{4}\cos^22\theta_W}{2}\,.
\end{equation}
The remaining parameters are
\begin{eqnarray}
\nonumber A&=& \frac{|a_R^Z|^2+|a_L^Z|^2}{2} L_+\,,\\
\nonumber B&=& \frac{m_t^2}{v^2}\frac{|b_{LR}^Z|^2+|b_{RL}^Z|^2}{2}  L_+\,,\\
\nonumber C&=&-\frac{m_t}{v}\frac{\mathrm{Re}\{b^Z_{LR}a_L^{Z*}+b_{RL}^Za_{R}^{Z*}\}}{2} L_+\,,\\
\nonumber  \alpha &=& \frac{|a_R^Z|^2-|a_L^Z|^2}{2}L_-\,,\\
\nonumber  \beta &=& \frac{m_t^2}{v^2}\frac{|b_{LR}^Z|^2-|b_{RL}^Z|^2}{2}L_-\,,\\
\nonumber \gamma&=&\frac{m_t}{v}\frac{ \mathrm{Re}\{b_{LR}^Za_L^{Z*}-b_{RL}^{Z}a_{R}^{Z*}\}}{2}L_-\,.
\label{Zconst}
\end{eqnarray}
We also use normalized mass and total decay width of $Z$ boson and define $\tilde{\Gamma}$
$$
\hat{m}_Z = \frac{m_Z^2}{m_t^2}\,,\hspace{0.5cm} \gamma_Z = \frac{\Gamma_Z}{m_Z}\,, \hspace{0.5cm} \tilde{\Gamma}=\frac{\Gamma }{\frac{m_t}{16\pi^3}\frac{g_Z^4v^4}{\Lambda^4}}\,.
$$
We can write the total decay rate as
\begin{eqnarray}
\Gamma &=&  \frac{m_t}{16\pi^3}\frac{g_Z^4v^4}{\Lambda^4}\Big[f_A A + f_B B + f_C C\Big]\,,
\end{eqnarray}
while the two asymmetries read
\begin{eqnarray}
A_{\mathrm{FB}} &=& \frac{1}{\tilde{\Gamma}}f_{\alpha\beta\gamma}[\alpha -4\beta+4\gamma]\,,\\
\nonumber A_{\mathrm{LR}} &=&\frac{ g_A A + g_B B + g_C C + g_{\alpha\beta\gamma}[\alpha-4\beta+4\gamma]}{\tilde{\Gamma}}\,.\\
\end{eqnarray}
The $f$ and $g$ functions depend only on the $Z$ boson parameters - mass and total decay width. They are presented in  
eqs.~(\ref{fA}-\ref{gZ}) of appendix A. We explore the possible ranges and correlations between the two asymmetries in fig.~\ref{fig:assym_Z} using
PDG stated values for evaluating the $f$ and $g$ functions
\begin{eqnarray*}
m_Z = 91.2\,  \mathrm{GeV}\,,\hspace{0.5cm} \Gamma_Z = 2.5\, \mathrm{GeV}\,,\\
m_t = 171.2\, \mathrm{GeV}\,,\hspace{0.5cm}\sin^2\theta_W = 0.2312\,.
\end{eqnarray*}
On the same plot we also project the limits, where only dipole or only current interactions of the  $Z$ contribute. In ref.~\cite{Fox:2007in} strong indirect limits were reported on the left-handed FCNC couplings of the $Z$ coming from low energy observables. Therefore we also superimpose the possible predictions for the two asymmetries when these couplings are set to zero.

The main difference between our notation and that of ref.~\cite{Fox:2007in} is that their manifest $SU(2)_L$ invariance relates selected FCNC operators 
containing left-handed top and bottom quarks; on the other hand we are only interested in top quark FCNC phenomenology warranting the more 
compact notation.
For completeness, the complete matching of our coupling constants to those of ref.~\cite{Fox:2007in} is presented in eqs.~(\ref{eq:fox1}-\ref{eq:fox6}) of appendix~B. 
We observe that the LRA can be used to distinguish between dipole and current FCNC couplings of the Z, while the FBA can distinguish the chiralities of the couplings.

Finally let us briefly comment on the application of our results to the warped extra dimensional Randal-Sundrum (RS) models. 
The authors of ref.~\cite{Agashe:2006wa} and ~\cite{Casagrande:2008hr} consider somewhat different implementations of the RS models and both find 
$tcZ$ couplings that could lead to observable $t\rightarrow c Z$ decays. While it may be tempting to apply our analysis to the two scenarios directly,
the results would not be significant. This is due to the fact that the two models could very well also have relevant $tc\gamma$ and $tch$ couplings
which would alter the $t\rightarrow cl^+l^-$ three body decay properties. We study such generic cases in the next sections.

\subsection{Interference of photon and $Z$ mediation}

\begin{figure}[h]
\begin{center}
\includegraphics[scale=1 ]{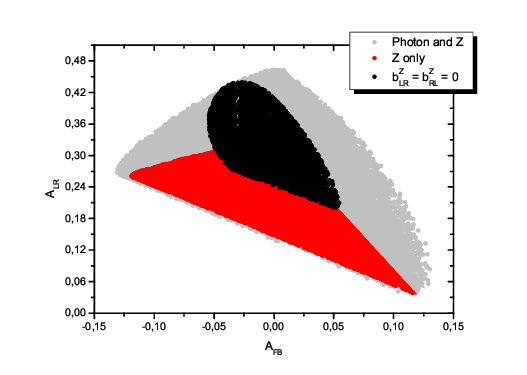}
\caption{\label{fig:assym_Zg}\small The correlation of FBA and LRA in $Z$ and $\gamma$ mediated decay. The gray area represents decays with all possible $Z$ and
photon FCNC couplings. The black area corresponds to decays with only current $Z$ FCNC couplings. For comparison, the red area represents $Z$ mediated decays.  }
\end{center}
\end{figure}

Several NP models predict comparable decay rates for $t\to c(u) Z,\gamma$. 
This may in turn lead to a situation, where an experimental search using a common 
final state may be more promising than dedicated  searches in each channel separately.
In addition, the asymmetries in $t\to c(u)\ell^+\ell^-$ may shed additional light on the specific couplings involved. 
The decay rate in this case depends again on the di-lepton invariant mass cutoff $\epsilon$
\begin{eqnarray}
\Gamma(\epsilon) =  \Gamma^{\gamma}(\epsilon) + \Gamma^{Z}(\epsilon) + \Gamma^{\mathrm{int}}(\epsilon)\,
\end{eqnarray}
where the last term represents the $\gamma$, $Z$  interference contribution.
\begin{eqnarray}
\Gamma^{Z} &=&\frac{m_t}{16\pi^3}\frac{v^4g_Z^4}{\Lambda^4}\Big[f_A^{\epsilon}A+f_B^{\epsilon}B+f_C^{\epsilon}C\Big]\,,\\
\nonumber \Gamma^{\mathrm{int}}&=&\frac{m_t}{16\pi^3}\frac{v^4g_Z^4}{\Lambda^4}\Big[f_{W_{12}} (W_1+W_2) +f_{W_{34}}( W_3+ W_4)\Big]\,.\\
\end{eqnarray}
Here $W_1,\dots,W_4$ are model dependent constants containing both $Z$ and $\gamma$ couplings 
\begin{eqnarray}
\nonumber W_1&=& \frac{m_t^2}{v^2}\frac{e^2}{g_Z^2}\frac{1}{2}\mathrm{Re}\{b_{LR}^{\gamma*}b_{LR}^{Z}c_L+b_{RL}^{\gamma*}b_{RL}^{Z}c_R \} \,, \\
\nonumber W_2&=& \frac{m_t^2}{v^2}\frac{e^2}{g_Z^2}\frac{1}{2}\mathrm{Re}\{b_{LR}^{\gamma*}b_{LR}^{Z}c_R+b_{RL}^{\gamma*}b_{RL}^{Z}c_L \}\,, \\
\nonumber W_3&=& \frac{m_t}{v}\frac{e^2}{g_Z^2}\frac{1}{2}\mathrm{Re}\{-b_{LR}^{\gamma*}a_L^{Z}c_L - b_{RL}^{\gamma*}a_{R}^{Z}c_R \}\,, \\
\nonumber W_4&=& \frac{m_t}{v}\frac{e^2}{g_Z^2}\frac{1}{2}\mathrm{Re}\{-b_{LR}^{\gamma*}a_L^{Z}c_R - b_{RL}^{\gamma*}a_{R}^{Z}c_L \}\,.
\end{eqnarray}
To shorten the notation we denote the SM lepton couplings to the $Z$ appearing in eq.~(\ref{eq:L-SM2}) by $c_R$ and $c_L$.
For the two asymmetries we get
\begin{eqnarray}
\nonumber A_{\mathrm{FB}} &=&  \frac{1}{\tilde{\Gamma}}\Big[f_{\alpha\beta\gamma}^{\epsilon} (\alpha-4\beta+4\gamma)+ \\
&&+ f_W \big(2(W_2-W_1)+ W_4-W_3\big)\Big]\,,\\
\nonumber A_{\mathrm{LR}} &=&  \frac{1}{\tilde{\Gamma}}\Big[g_A^{\epsilon} A + g_B^{\epsilon} B+g_C^{\epsilon} C+ \\
\nonumber &&+g_{\alpha\beta\gamma}(\alpha-4\beta+4\gamma)+\\
&&+\sum_{i=1}^4 g_{W_i} W_i + g_{\gamma}B_{\gamma}\Big]\,.
\end{eqnarray}
The functions $f$ and $g$ depend now on the $Z$ boson parameters and also the di-lepton invariant mass cutoff $\epsilon$. They are presented in 
eqs.~(\ref{fW12}-\ref{gZg}) of appendix A.
We plot the possible correlation between the FBA and the LRA in this scenario with a fixed cut on $s$ set to $40$ GeV in fig.~\ref{fig:assym_Zg}.
We also present possible points for the case when only the current FCNC $Z$ couplings contribute.
We observe that in principle interference effects can produce a larger LRA compared to the case of pure $Z$ mediation.

In ref.~\cite{Fox:2007in} upper bounds on coefficients accompanying the operators responsible for FCNC $t\to c Z$ and $t\to c \gamma$ are presented. Using transcription formulae presented in eqs. (\ref{eq:fox1}-\ref{eq:fox6}) of appendix B we can evaluate AFB and ALR associated with these upper bounds. The numerical values are presented in the table-\ref{tabela}.
\begin{center}
\begin{table}
\begin{tabular}{|c|c|c|}\hline
 & only $Z$ & $Z$ and $\gamma$ \\\hline
$A_{\mathrm{FB}}$ &0.045 & 0.035\\
$A_{\mathrm{LR}}$ &0.206 & 0.226\\\hline
\end{tabular}
\caption{\small Values of AFB and ALR for highest allowed coefficients given by Fox et al. in ref.~\cite{Fox:2007in}. The VEV $v$ is set to $v = 174$ GeV. The cutoff $\epsilon$ for the second column is again taken to be $40$ GeV.}
\label{tabela}
\end{table}
\end{center}
These values serve just for illustration that nonzero values of asymmetries can indeed be obtained. They do not represent any
kind of upper bounds for asymmetries. There is no reason to think that the highest allowed values of coefficients (\ref{eq:fox1}-\ref{eq:fox6})
are to give the largest possible asymmetries which are complicated functions of these coefficients.

\subsection{Light Higgs mediation}
\label{SMHiggs}

Considering a light Higgs coupling to leptons, a major difference with the previous cases is the large dependence on the lepton flavour. It is only expected to contribute significantly to the mode with tau leptons in the final state, making its detection quite challenging (the primary strategy for detection of $t c \phi$ couplings is currently via the $b\bar b$ decay mode of the Higgs). However we consider it here for completeness, since it also applies to the case of possible BSM scalar resonances below the top mass. Again we denote

$$
C_h =\frac{m_t^2}{v^2}\frac{1}{g_Z^4}\frac{|c_{LR}^h|^2+|c_{RL}^h|^2}{2}\,,\hspace{0.5cm} \hat{m}_{\phi} = \frac{m_{\phi}^2}{m_t^2}\,,\hspace{0.5cm} \gamma_{\phi}=\frac{\Gamma_{\phi}}{m_{\phi}}\,.
$$
in terms of which 
\begin{equation}
\Gamma = \frac{m_t}{16 \pi^3}\frac{g_Z^4v^4}{\Lambda^4}\hat{m}_l C_{h} f_h\,.
\end{equation}
Again the FBA vanishes identically, while for the LRA can be written as
\begin{equation}ž
A_{\mathrm{LR}} = \frac{g_h}{f_h}\,.
\end{equation} 
where $g_h$ and $f_h$ depend only on the Higgs boson parameters. We plot this dependence in fig.~\ref{fig:alr_h} while the analytic expressions are
presented in eqs.~(\ref{fh}, \ref{gh}) of appendix A.
\begin{figure}[h]
\begin{center}
\includegraphics[scale=1]{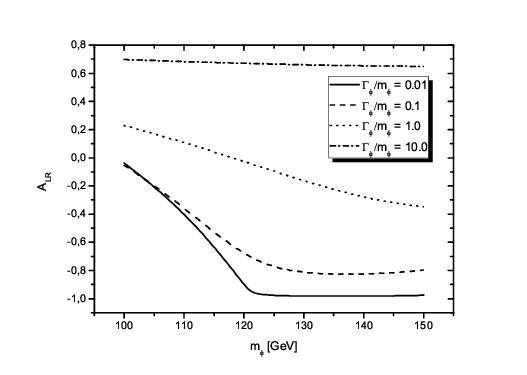}
\caption{\label{fig:alr_h}\small Dependence of the LRA in light Higgs mediated transition on the Higgs parameters.}
\end{center}
\end{figure}
We see that in the case of light Higgs mediation, the LRA does not depend on the coupling parameters, but is strongly sensitive to the Higgs mass and the total decay width.
For small $\Gamma_{\phi}/m_{\phi}$ ratios the LRA is strictly negative on the presented mass interval, but for larger ratios it
can also be positive.

The above discussion applies to the two-Higgs-doublet models for the top.
In ref.~\cite{Baum:2008qm} the authors predict $c_{RL} = 1$ and $c_{LR}=\epsilon_{ct}=m_c/m_t \approx 7.3\cdot 10^{-3}$ in the Higgs mediated mode.
However one would possibly need to measure spin-related observables associated with the polarization of the final state leptons (taus) in order to disentangle the two contributions. 
The authors of ref.~\cite{Gaitan:2004by} also predict observable branching ratios for $t\rightarrow c h$ and $t\rightarrow c Z$ in the framework of Alternative Left-Right Symmetric models. We study possible interference effects of these modes in the next section.

\subsection{Interference between light Higgs and $Z$ mediation}

Let us note that any possible interference between scalar and vector resonance contributions in $t\to c(u) \ell^+\ell^-$ are necessarily chirally suppressed by the light fermion masses. 
However the same holds for the Higgs coupling themselves. We therefore do not neglect the final state masses 
when considering the matrix element of the interference term.
The decay rate is
$$
\Gamma=\Gamma^{Z}+\Gamma^{\phi}+\Gamma^{\mathrm{int}}\,
$$
where the last term is the interference contribution. 
The model parameters are
\begin{eqnarray}
\nonumber E &=& \frac{1}{4}\frac{m_t}{v}\frac{1}{g_Z^2}(c_L+c_R)(a_R^Z c_{LR}^{h*}+a_L^Z c_{RL}^{h*})\,,\\
\nonumber F &=& \frac{1}{4}\frac{m_t^2}{v^2}\frac{1}{g_Z^2}(c_L+c_R)(b_{LR}^Z c_{LR}^{h*}+b^Z_{RL}c_{RL}^{h*})\,,
\end{eqnarray}
where $c_L$ and $c_R$ are again the SM lepton couplings to the $Z$ boson from eq.~(\ref{eq:L-SM2}).
As we integrate $\dd \Gamma^{\mathrm{int}}/(\dd \uy\dd\sy)$ over $\uy$ we find $\dd \Gamma^{\mathrm{int}}/\dd \sy = 0$. 
The interference term does however contribute to both asymmetries. 
\begin{eqnarray}
\nonumber\Gamma &=& \frac{m_t}{16\pi^3}\frac{g_Z^4v^4}{\Lambda^4}\times \\
 && \Big[f_A A+f_B B+f_C C+ \hat{m}_lf_h C_h\Big],\\
\nonumber A_{\mathrm{FB}} &=&\frac{1}{\tilde{\Gamma}}\Big[f_{\alpha\beta\gamma}(\alpha-4\beta+4\gamma)+\\
&&+\hat{m}_l(f_E E+f_F F)\Big]\,,\\
\nonumber A_{\mathrm{LR}} &=&\frac{1}{\tilde{\Gamma}}\Big[g_A A+g_B B+g_C C+\\
\nonumber &&+ g_{\alpha\beta\gamma}(\alpha-4\beta+4\gamma)+\\
&&+ \hat{m}_l(g_h C_h+ g_E E+g_F F)\Big]\,.
\end{eqnarray}
The new functions $f_{E,F}$ and $g_{E,F}$ depend on both $Z$ and Higgs parameters.
They are presented in eqs.~(\ref{fE}-\ref{gEF}) of appendix A.
\begin{figure}[h]
\begin{center}
\includegraphics[scale=1 ]{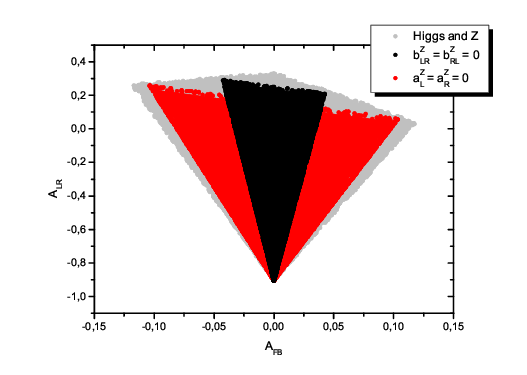}
\caption{\label{fig:assym_Z_H1}\small The correlation of FBA and LRA in $Z$ and light Higgs mediated decay. The gray area represents decays with all possible $Z$ and light Higgs FCNC couplings, while the black and red areas correspond to decays with only current and only dipole $Z$ FCNC couplings.}
\end{center}
\end{figure}
\begin{figure}[h]
\begin{center}
\includegraphics[scale=1 ]{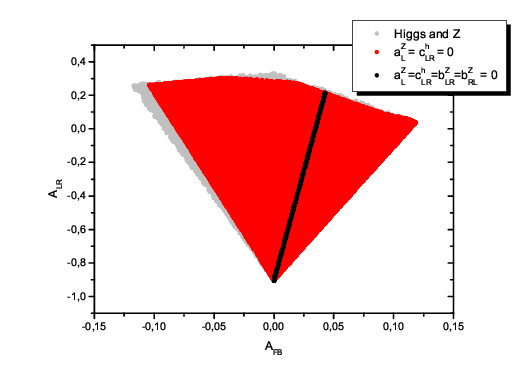}
\caption{\label{fig:assym_Z_H2}\small The correlation of FBA and LRA in $Z$ and light Higgs mediated decay. The gray area represents decays with all possible $Z$ and light Higgs FCNC couplings, while the red area corresponds to decays with $a_L^Z$ and $c_{LR}^h$ set to zero. The black line represents the decays where in addition, the dipole $Z$ FCNC couplings are also set to zero. }
\end{center}
\end{figure}
In fig.~\ref{fig:assym_Z_H1} and fig.~\ref{fig:assym_Z_H2} we present possible correlations between the FBA and the LRA.
The scalar particle parameters are set to $m_{\phi}=120$ GeV, which is just above the PDG stated lower limit for a SM-like Higgs boson, and $\Gamma_{\phi}=1$ GeV. For the lepton pair
we choose $\tau^+\tau^-$ with $m_{\tau}= 1.8$ GeV.
In fig.~\ref{fig:assym_Z_H1} we plot, in addition to the points obtained when all the $Z$ and Higgs FCNC couplings are considered, points where only the current
or only the dipole FCNC $Z$ couplings contribute. 
In fig.~\ref{fig:assym_Z_H2} we take into account the limits of ref.~\cite{Fox:2007in} and set $a_L^Z=c^h_{LR}=0$.

We can see that the region of possible $(A_{\mathrm{FB}},A_{\mathrm{LR}})$ points expands noticeably in comparison to the $Z$ only mediated decay. In contrast to the expansion that occurs when we consider $Z$ and photon interference, the area now expands to lower LRA values, also allowing for negative LRA.

\subsection{Heavy vector ($Z'$) $s$-channel exchange}

\begin{figure}[h]
\begin{center}
\includegraphics[scale=1]{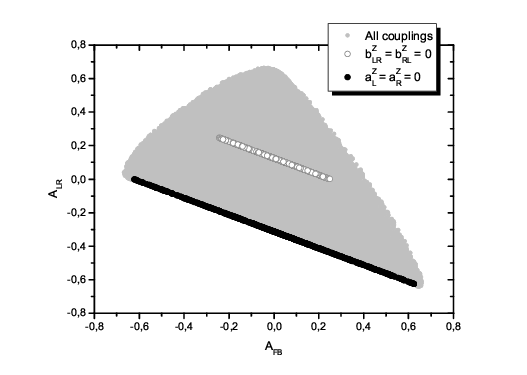}
\caption{\label{fig:assym_Zp}\small The correlation of FBA and LRA in the $Z'$ mediated decay. The gray area represents decays with all possible current and dipole $Z'$ FCNC couplings. The white and black lines represent decays with only current and only dipole couplings respectively.}
\end{center}
\end{figure}

This case is very similar to the $Z$ mediation. However, results from direct searches at LEP and Tevatron for vector resonances decaying into pairs of leptons put severe bounds on possible $Z'$ mass $m_{Z'}\gtrsim 1$~TeV. At these values, we can completely neglect the $Z'$ width effects as well as any $s$ dependence in the denominator of the amplitudes. Apart from this, we are using the notation presented in (\ref{Zconst}), only now the leptonic couplings to $Z'$ need now no longer be SM-like, so $L_+$ and $L_-$ change to
\begin{equation}
\nonumber L_{\pm}= \frac{|a^l_R|^2\pm|a^l_L|^2}{2}\,.
\end{equation}
For the total rate we thus obtain
\begin{eqnarray}
\Gamma &=&\frac{m_t}{16\pi^3} \frac{g_Z^4v^4}{\Lambda^4} \frac{1}{\hat{m}_{Z'}^2}\frac{1}{120}\Big[5 A + 8 B + 10C \Big]\,,
\end{eqnarray}
while the two asymmetries read
\begin{eqnarray}
A_{\mathrm{FB}} &=& \frac{-\frac{5}{4}\alpha +5\beta-5\gamma }{5 A + 8 B + 10 C}\,,\\
A_{\mathrm{LR}}&=&\frac{\frac{5}{4}\frac{A+\alpha}{2}-5\frac{B+\beta}{2}-5\frac{C-\gamma}{2}}{5A+8B+10C}\,.
\end{eqnarray}
We explore the possible ranges and correlations between the two asymmetries in fig.~\ref{fig:assym_Zp}. 
On the same plot we also project the limits, where only the dipole or only the current interactions of the  $Z'$ contribute. We note that these results also apply to (box) loop induced~\footnote{Here we must assume the particles entering the loop are much heavier than the top, so we can neglect the dependence on the external momenta.} or effective contact interactions which can be written as the product of quark and leptonic currents. Lastly we note that charged baryon and lepton number violating vector resonance exchange (or corresponding effective contact interactions) can be accommodated via the interchange of the two asymmetries.

\subsection{Heavy scalar ($\phi'$)}

\begin{figure}[h]
\begin{center}
\includegraphics[scale=1]{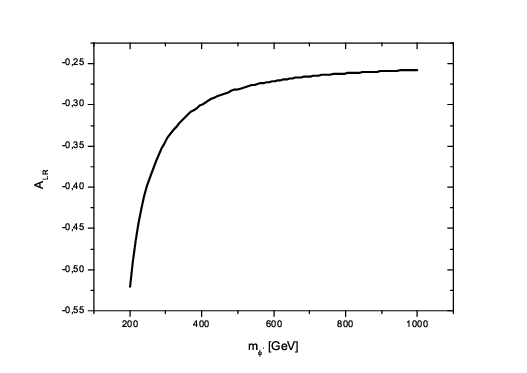}
\caption{\label{fig:alr_hh}\small Dependence of the LRA in scalar mediated decay on the scalar resonance mass.}
\end{center}
\end{figure}

Again this case is similar to the light Higgs mediation, where now we consider arbitrary scalar lepton couplings and also neglect width effects in the heavy scalar propagators. The notation is now
$$
C_{h'} = \frac{|c^{h'}_{LR}|^2+|c^{h'}_{RL}|^2}{2}\,,\hspace{0.5cm}C_l = \frac{|c^{l}_{LR}|^2+|c^{l}_{RL}|^2}{2}
$$
yielding for the total rate
\begin{eqnarray}
\Gamma &=&  \frac{m_t}{128 \pi^3}\frac{v^4}{\Lambda^4}C_{h'}C_l f_{\phi'}
\end{eqnarray}
and for the LRA (the FBA is of course zero)
\begin{eqnarray}
A_{\mathrm{LR}} = \frac{g_{\phi'}}{f_{\phi'}}\,.
\end{eqnarray}
The LRA is independent of the couplings as functions $f$ and $g$ in this case only depend on the scalar mass as 
presented in eqs.~(\ref{fhprime}, \ref{ghprime}) of appendix A.
LRA is presented as the function of the scalar mass in fig.~\ref{fig:alr_hh}.
We could at this point consider interference between heavy vector and heavy scalar mediated decays.
In general, when $\phi'$ couplings to fermions are not proportional to 
their masses as in the case of the SM Higgs \footnote{Typically this is the case when the relevant scalar field does not 
develop a vacuum expectation value (e.g. squarks in MSSM with RPV) 
or its VEV gives subdominant contributions to fermion masses.}, there is no interference 
contribution to be considered in the ultrarelativistic limit, so that the $Z'$ and $\phi'$ contributions can be
summed incoherently.


\section{Conclusions}

We have considered the FCNC top quark decay modes $t\to c (u)\ell^+\ell^-$ as a probe of BSM physics at the LHC. In addition to the branching ratio, we have defined two angular asymmetries which can serve to further discriminate between different NP scenarios. Comparing all possible contributions to the decay mode via SM field mediation as well as BSM resonance exchange in both $s$- and $u$-channel corresponding also to effective contact interactions, we can draw the following general conclusions: large values of FBA ($|A_{\mathrm{FB}}|\gg 0.1$) cannot be accounted for in decay modes mediated by SM bosons as long as we assume these bosons to have SM couplings to the charged leptons. We have shown in fig.~\ref{fig:assym_Z}, that the $A_{\mathrm{FB}}\in[-0.12,0.12]$. Larger values of FBA, $A_{\mathrm{FB}}\in[-0.66,0.66]$, could appear in $Z'$ mediated decays or in $u$-channel exchange, where FBA and LRA exchange their roles. 
This could signal the presence of baryon and lepton number violating interactions~\footnote{Alternatively, loop induced contributions with light particles in the loop could induce large dependence on the external momenta and also produce sizable asymmetries.}. On the other hand large negative values of LRA ($A_{\mathrm{LR}} \ll -0.2$) and vanishing FBA could indicate a contribution due to a relatively light scalar. 
A measured point in $(A_{\mathrm{FB}},A_{\mathrm{LR}})$ plane could exclude models with only current or only dipole FCNC couplings of $Z$ or $Z'$ if it were located off the white or black lines in fig.~\ref{fig:assym_Z} and ~\ref{fig:assym_Zp}.
Treating the $Z$ and photon or the $Z$ and light Higgs mediated decays as indistinguishable expands the allowed LRA region. In the first case to larger positive values and in the later to smaller and even negative values of the LRA.

Current experimental sensitivity studies look at the two body decay modes $t\to c Z,\gamma$ ~\cite{Carvalho:2007yi}.

Our analysis may be applicable to the potential measurement of $t\to c Z$ at ATLAS since they will be identifying the Z boson through its decay to a lepton pair.
Angular asymmetries of this pair and the remaining hard jet could provide additional information on the $tcZ$ FCNC vertex. The
$t\to c\gamma$ decay is generically characterized by a single high $p_T$ photon. Current search strategies for this FCNC include the detection of this photon, and not its eventual 
decay to a lepton pair.
FCNC top quark decays mediated by heavy vector or scalar bosons are at present not being considered by ATLAS or CMS. These new
heavy particles will be searched for as narrow resonances decaying into dilepton pairs~\cite{Aad:2009wy}. 

In order to fully explore our decay mode, one would need to relax or modify certain criteria used by current search strategies to reduce SM backgrounds. In addition, the reconstruction of the LRA might require top quark charge tagging. The resulting loss of sensitivity could be perhaps compensated using new techniques for jet ~\cite{Butterworth:2008iy,Butterworth:2008tr} and tau lepton identification~\cite{Friis:2008tg} (relevant for Higgs channels). Especially the later might provide new additional interesting signatures by leveraging spin self-analyzing properties of taus. In principle our results are applicable also to the purely hadronic decay modes, where the two leptons are replaced by $b$-tagged jets for example, however in this case the asymmetries are compromised by the lack of knowledge of the sign of the $b$-quark charges.

The present work could be extended in several directions. As already mentioned above, precise study of backgrounds and strategies to discriminate against them would be crucial to properly evaluate the potential of the proposed observables. Related to this, a proper simulation of jet formation, relevant experimental cuts and detector effects in our decay mode are in progress and will be presented elsewhere. 
In principle we expect the measurement strategies (and their limitations) for these asymmetries to be similar to those already devised~\cite{Aad:2009wy} for the helicity structure of the $tbW$ couplings, as already successfully measured at the Tevatron~\cite{Abulencia:2006ei, Abazov:2007ve, Aaltonen:2008ei}, with the added benefit that the event kinematics in our case can be fully reconstructed.
Also, $\alpha_s$ corrections to the considered observables could be important~\cite{Zhang:2008yn}. In particular mixing with the chromomagnetic FCNC operator is expected~\cite{Ferreira:2008cj} at NLO which could enhance the predicted rates.

Finally the $e^+ e^- \to t \bar c$ cross-section is related to our decay mode via crossing symmetry. The relevance of this mode at the next linear collider (NLC) has already been considered in specific models~\cite{Wang:2008qn, Agashe:2006wa, Chen:2008za}. In this case however, due to the fixed center-of-mass energy, only single asymmetry can be defined. On the other hand, the possibility of polarized beams~\cite{AguilarSaavedra:2004wm} gives access to new spin and CP observables~\cite{Antipin:2008zx}.

\begin{acknowledgments}
We are very thankful B. P. Ker\v sevan for useful discussions and Ulrich Husemann for valuable comments.
This work is supported in part by the EU-RTN Programme, Contract No. MRTN--CT-2006-035482, \lq\lq Flavianet'' and by the Slovenian Research Agency.
\end{acknowledgments}

\appendix


\section {Analytic formulae}
Below we give the complete analytic formulae for the partial differential decay rate distributions
in terms of our chosen kinematical variables. With the substitution to angular variables and after integration the FBA and LRA
can be obtained from these expressions. The notation adheres to the conventions set in eqs.~(1-4). We
also present the expressions for the $f$ and $g$ functions appearing in the text. Mostly they are given 
in unevaluated integral form, as analytic integration, though possible in most cases, yields very long expressions.
\subsection{Photon mediation}

\begin{eqnarray}
\nonumber \frac{\dd\Gamma}{\dd \uy\dd\sy}&=&\frac{m_t}{16\pi^3}\frac{g_Z^4v^4}{\Lambda^4}B_{\gamma}\times\\
 &&\frac{1}{\sy}\Big[\sy(2\uy-1) + 2\uy^2-2\uy+1\Big]\,,\\
\frac{\dd\Gamma}{\dd\hat s} &=& \frac{m_t}{16\pi^3}\frac{g_Z^4v^4}{\Lambda^4} B_{\gamma} \frac{(1-\sy)^2(\sy+2)}{3\sy}\,.
\end{eqnarray}

\begin{eqnarray}
f_{\gamma} &=& \frac{1}{9} \Big[-\epsilon^3 + 9\epsilon-6\log(\epsilon)-8\Big]\,,\label{fg}\\
g_{\gamma} &=&-\frac{13}{18}+3\epsilon-2\epsilon^2+\epsilon^3-\frac{2}{3}\log(4\epsilon)\,.\label{gg}
\end{eqnarray}

\subsection{$Z$ mediation}

\begin{eqnarray}
\nonumber \frac{\dd \Gamma}{\dd \sy\dd \uy}&=&\frac{m_t}{16\pi^3}\frac{g_Z^4v^4}{\Lambda^4}\frac{1}{(\sy-\hat{m}_Z)^2+\sy^2\gamma_Z^2}\times\\
\nonumber &&\Big[\frac{ A + \alpha}{4}(1- \sy - \uy )( \sy + \uy) +\\
\nonumber &&+\frac{ A - \alpha}{4} (1- \uy)\uy  + \\
\nonumber &&+(B+\beta)\uy\sy(\uy+\sy)+ \\
\nonumber &&+(B-\beta)\sy (1-\sy-\uy) (1-\uy)+\\
&&+(C+\gamma)\sy (1-\uy-\sy)+(C-\gamma)\uy\sy\Big]\,,\\
\nonumber \frac{\dd\Gamma}{\dd\hat s} &=&  \frac{m_t}{16\pi^3}\frac{g_Z^4v^4}{\Lambda^4}\frac{(\sy-1)^2}{(\sy-\hat{m}_Z)^2 + \sy^2\gamma_Z^2}\times\\
&&\Big[\frac{A}{12}(2\sy+1) + \frac{B}{3}\sy(\sy+2) + C \sy\Big]\,.
\end{eqnarray}
To shorten the notation we define
\begin{eqnarray}
\nonumber r_1 &=& \frac{(1-\sy)^2}{(\sy - \hat{m}_Z)^2 + \sy^2\gamma_Z^2}\,,\\
\nonumber r_2 &=& \frac{\frac{1}{8}(1-\uy)^2}{[(1-z)(1-\uy)-2\hat{m}_Z]^2 + \gamma_Z^2(1-z)^2(1-\uy)^2	}\,.
\end{eqnarray}
\begin{eqnarray}
f_A&=&\int_{0}^{1} \dd \sy \,r_1\, \frac{1}{12}(1+2\sy)\,,\label{fA}\\
f_B&=&\int_{0}^{1} \dd \sy \,r_1\, \frac{1}{3}(2\sy+\sy^2)\,,\label{fB}\\
f_C&=&\int_{0}^{1} \dd \sy \,r_1\, \sy\,,\label{fC}\\
f_{\alpha\beta\gamma} &=& -\frac{1}{8} f_C\,.\label{fabc}
\end{eqnarray}
The $g$ functions present in LRA expressions are more complicated due to the fact that 
the angular variable appears in the resonant factor of the matrix element. So for the sake of brevity
we define additional functions $G$ in which the $\uy$ integration is performed.
\begin{eqnarray}
g_X&=&\int_0^1 \dd z \,G_X - \int_{-1}^{0}\dd z  \,G_X\,,\label{gZ}\\
\nonumber G_A &=&\int_{0}^{1} \dd \uy\, r_2\, (1+5\uy+2\uy z-z^2+\uy z^2)\,,\\
\nonumber G_B &=&\int_{0}^{1} \dd \uy\, r_2\, 4(1-\uy+2\uy^2-\\
\nonumber &&-2\uy z-z^2+3\uy z^2-2\uy^2z^2)\,,\\
\nonumber G_C&=&\int_{0}^{1} \dd \uy\, r_2\, 4(1 + \uy - 2\uy z - z^2 +\uy z^2)\,,\\ 
\nonumber G_{\alpha\beta\gamma} &=&\int_{0}^{1} \dd \uy\, r_2\, (1 - 3 \uy + 2\uy  z - z^2 + \uy  z^2)\,.
\end{eqnarray}


\subsection{Interference between $Z$ and photon mediation}

The interference contribution between the $Z$ and the photon to the double-differential decay rate is
\begin{eqnarray}
\nonumber\frac{\dd \Gamma^{\mathrm{int}}}{\dd \sy\dd\uy}&=&\frac{m_t}{16\pi^3}\frac{v^4g_Z^4}{\Lambda^4}\mathrm{Re}\Bigg\{\frac{\sy-\hat{m}_Z-\ii\sy\gamma_Z}{(\sy-\hat{m}_Z)^2+\sy^2\gamma_Z^2}\times\\
\nonumber &&\Big[2W_1(1-\sy-\uy)(1-\uy)+2W_2\uy(\uy+\sy)+\\
 &&+W_3(1-\sy-\uy) + W_4\uy\Big]\Bigg\}\,.
\end{eqnarray}
In all further computations we neglect the imaginary part in the propagator's numerator 
$$
\gamma_Z \sim 0.02 \Rightarrow \gamma_Z\ll 1\,.
$$
This means that $\mathrm{Re}$ acts only on the model dependent constants $W_1,\dots,W_4$.
$f_X^{\epsilon}$ and $g_X^{\epsilon}$ are the same as $f_X$ and $g_X$, except
that the integration limits are altered due to the di-lepton invariant mass cutoff $\epsilon$.
In $f_X$ the $\sy$ integration is now in the $[\epsilon,1]$ region, in $g_X$ the 
intervals for $z$ are $[0,1-2\epsilon]$ and $[-1,0]$, and for the $\uy$ in $G_X$ functions
$\uy\in[0,1-\frac{2\epsilon}{1-z}]$. 
We further define
\begin{eqnarray}
\nonumber r_3 &=& \frac{[(1-\uy)(1-z)-2\hat{m}_Z](1-\uy)}{[(1-z)(1-u)-2\hat{m}_Z]^2 + \gamma_Z^2(1-z)^2(1-u)^2}\,.
\end{eqnarray}
The new $f$ and $g$ functions are
\begin{eqnarray}
f_{W_{12}} &=&\int_{\epsilon}^1 \dd \sy\, r_1\,(s-\hat{m}_Z) \frac{1}{3} (\sy +2)\,,\label{fW12}\\
f_{W_{34}} &=&\int_{\epsilon}^1 \dd \sy\, r_1\,(s-\hat{m}_Z) \frac{1}{2}\,,\label{fW34}\\
f_{W} &=&\frac{1}{2} f_{W_{34}}\,,\label{fW}\\
g_X&=&\int_0^{1-2\epsilon} \dd z \,G_X - \int_{-1}^{0}\dd z  \,G_X\,,\label{gZg}\\
\nonumber G_{W_1}&=&\int_{0}^{1-\frac{2\epsilon}{1-z}} \dd \uy\, r_3\, (1-\uy)^2(1+z)\,,\\ 
\nonumber G_{W_2}&=&\int_{0}^{1-\frac{2\epsilon}{1-z}} \dd \uy\, r_3\, \uy (1+\uy-z+z\uy)\,,\\ 
\nonumber G_{W_3}&=&\int_{0}^{1-\frac{2\epsilon}{1-z}} \dd \uy\, r_3\, \frac{1}{2} (1+\uy+z-z\uy)\,,\\ 
\nonumber G_{W_4}&=&\int_{0}^{1-\frac{2\epsilon}{1-z}} \dd \uy\, r_3\, \uy \,.
\end{eqnarray}

\subsection{Light Higgs mediation}

\begin{eqnarray}
\frac{\dd \Gamma}{\dd \sy \dd \uy} = \frac{m_t}{16 \pi^3}\frac{v^4g_Z^4}{\Lambda^4}\frac{1}{8}\hat{m}_l C_h \frac{\sy (1-\sy)}{(\sy-\hat{m}_{\phi})^2-\sy^2\gamma_{\phi}^2}\,,
\end{eqnarray}
\begin{eqnarray}
f_h &=& \int_0^1\dd\sy\,\frac{1}{8} \frac{(1-\sy)^2\sy}{(\sy-\hat{m}_{\phi})^2+\sy^2\gamma_{\phi}^2}\,,\label{fh}\\
g_h&=&\int_0^{1} \dd z \,G_h - \int_{-1}^{0}\dd z  \,G_h\,,\label{gh}\\
\nonumber G_h &=& \int_0^1\dd\uy\, r_2\, \frac{1}{2}(z-1)(u(z-1)-z-1)\,.
\end{eqnarray}
The resonant part $r_2$ is the same as in the $Z$ mediated case, only the $Z$ parameters are substituted by
those of the light Higgs.

\subsection{Interference between $Z$ and Light Higgs mediation}

The interference contribution between the $Z$ and a light Higgs to the double-differential decay rate is
\begin{eqnarray}
\nonumber \frac{\dd \Gamma^{\mathrm{int}}}{\dd \sy \dd \uy} &=&\frac{m_t}{16\pi^3}\frac{g_Z^4v^4}{\Lambda^4}\hat{m}_l\frac{1}{2}(2\uy+\sy-1)\times\\
\nonumber &&\mathrm{Re}\Bigg\{\frac{(\sy-\hat{m}_{Z}-\ii \sy\gamma_Z)(\sy-\hat{m}_{\phi}+\ii \sy\gamma_{\phi})}
{[(\sy-\hat{m}_Z)^2+\sy^2\gamma^2_Z][(\sy-\hat{m}_{\phi})^2+\sy^2\gamma^2_{\phi}]}\times\\
&&\Big[E - 2\,F\,\sy\Big]\Bigg\}\,.
\end{eqnarray}
If we assume the $Z$ and the light Higgs FCNC couplings to be real, we need to consider only the
real part of the propagator's numerator. To shorten the notation we again define
\begin{eqnarray}
\nonumber r_4 &=& \frac{(\sy-\hat{m}_Z)(\sy-\hat{m}_{\phi})+\sy^2\gamma_{Z}\gamma_{\phi}}{[(\sy-\hat{m}_Z)^2+\sy^2\gamma_Z^2][(\sy-\hat{m}_{\phi})^2+\sy^2\gamma_{\phi}^2]}\,,\\
\nonumber r_5 &=& r_4[\sy\rightarrow \frac{1}{2}(1-\uy)(1-z)]\times \\
\nonumber &&\frac{1}{8}(1-\uy)(\uy(3+z)-z-1)\,.
\end{eqnarray}
New $f$ and $g$ functions are
\begin{eqnarray}
f_E &=& \int_0^1\dd \sy\,r_4\, \frac{1}{4}(1-\sy)^2\,,\label{fE}\\
f_F &=& \int_0^1\dd \sy\,r_4\, (-1)\frac{\sy}{2}(1-\sy)^2\,,\label{fF}\\
g_{E,F}&=&\int_0^{1} \dd z \,G_{E,F} - \int_{-1}^{0}\dd z  \,G_{E,F}\,,\label{gEF}\\
\nonumber G_E&=& \int_0^1\dd \uy\,r_5 \,,\\
\nonumber G_F&=& \int_0^1\dd \uy\,r_5\, (1-u)(z-1)\,.
\end{eqnarray}

\subsection{Heavy vector ($Z'$) mediation}

\begin{eqnarray}
\nonumber \frac{\dd\Gamma}{\dd\uy \dd\sy} &=& \frac{m_t}{64\pi^3}\frac{g_Z^4v^4}{\Lambda^4}\frac{1}{\hat{m}_{Z'}}\times \\
\nonumber &&\Big[(A+\alpha)(\uy+\sy)(1-\uy-\sy)+\\
\nonumber &&+ (A-\alpha)\uy(1-\uy)+\\
\nonumber &&+4(B+\beta)\uy\sy(\uy+\sy)+\\
\nonumber &&+4(B-\beta)\sy(1-\uy)(1-\sy-\uy)+\\
\nonumber &&+4(C+\gamma)\sy(1-\uy-\sy)+\\
&&+4(C-\gamma)\uy\sy\Big]\,,\\
\nonumber \frac{\dd\Gamma}{\dd\hat s} &=&  \frac{m_t}{16\pi^3}\frac{g_Z^4v^4}{\Lambda^4}\frac{(\sy-1)^2}{\hat{m}_{Z'}^2}\times\\
&&\Big[\frac{A}{12}(2\sy+1) + \frac{B}{3}\sy(\sy+2) + C \sy\Big]\,.
\end{eqnarray}
In the case of a $Z'$ mediated decay there is no need to define the $f$ and $g$ functions since the complete analytic expressions are simple enough
to be presented in the text.

\subsection{Heavy scalar ($\phi'$) mediation}

We do not assume the heavy scalar mass to necessarily be much greater than the top quark mass, so we do not neglect the $\sy$ in the propagator.
This gives us 
\begin{eqnarray}
\frac{\dd \Gamma}{\dd \sy\dd\uy}&=&\frac{m_t}{128\pi^3}\frac{v^4}{\Lambda^4}C_{h'}C_l \frac{\sy(1-\sy)}{(\sy-\hat{m}_{\phi'})^2}\,,\\
\frac{\dd\Gamma}{\dd\hat s} &=& \frac{m_t}{128 \pi^3}\frac{v^4}{\Lambda^4}C_{h'}C_l\frac{\sy(1-\sy)^2}{(\sy-\hat{m}_{\phi'})^2} \,.
\end{eqnarray}
Because we are not dealing with a resonant form of the propagator, we can present the $f$ and $g$ functions in their final analytic form
\begin{eqnarray}
\nonumber f_{\phi'}&=&-\frac{5}{2}+3\hat{m}_{\phi'}+\\
&&+(3\hat{m}_{\phi'}^2-4\hat{m}_{\phi'}+1)\log\Big[\frac{\hat{m}_{\phi'}-1}{\hat{m}_{\phi'}}\Big]\,,\label{fhprime}\\
\nonumber g_{\phi'}&=&(12\hat{m}_{\phi'}^2-12\hat{m}_{\phi'}+2)\log\Big[1-\frac{1}{2\hat{m}_{\phi'}}\Big]-\\
\nonumber &&-(3\hat{m}_{\phi'}^2-4\hat{m}_{\phi'}+1)\log\Big[1-\frac{1}{\hat{m}_{\phi'}}\Big]+\\
&&-2+3\hat{m}_{\phi'} \,.\label{ghprime}
\end{eqnarray}

\section{Matching to the parametrization of Fox et al.~\cite{Fox:2007in}}
Here we present the conversion of $\Leff$ presented in~\cite{Fox:2007in} to the form in eq.~(\ref{eq:L-SM}).
Fox et al. give a complete set of dimension six operators that give a $tcZ$ or  $tc\gamma$ vertex
\begin{eqnarray*}
O^u_{LL} &=&\ii \Big[\bar{Q}_3\tilde{H}\Big] \Big[(\slashed{D}\tilde{H})^{\dagger}Q_2\Big]
-\ii\Big[\bar{Q}_3(\slashed{D}\tilde{H})\Big] \Big[\tilde{H}^{\dagger}Q_2\Big] + \mathrm{h.c.}\,,\\
O_{LL}^h &=& \ii \Big[\bar{Q}_3\gamma^{\mu}Q_2\Big]\Big[H^{\dagger} (D_{\mu}H) - (D_{\mu}H)^{\dagger} H \Big] + \mathrm{h.c.}\,,\\
O_{RL}^w &=&g_2\Big[\bar{Q}_2\sigma^{\mu\nu}\sigma^a \tilde{H}\Big]t_R W^a_{\mu\nu} + \mathrm{h.c.}\,,\\
O_{RL}^b &=& g_1\Big[\bar{Q}_2\sigma^{\mu\nu}\tilde{H}\Big]t_R B_{\mu\nu} + \mathrm{h.c.}\,,\\
O_{LR}^w & =& g_2\Big[\bar{Q}_3\sigma^{\mu\nu}\sigma^a \tilde{H}\Big]c_R W_{\mu\nu}^a + \mathrm{h.c.}\,,\\
O_{LR}^b &=& g_1\Big[\bar{Q}_3\sigma^{\mu\nu} \tilde{H} \Big]c_R B_{\mu\nu} + \mathrm{h.c.}\,,\\
O^u_{RR} &=&\ii \bar{t}_R\gamma^{\mu}c_R \Big[H^{\dagger} (D_{\mu}H) - (D_{\mu}H)^{\dagger} H \Big] + \mathrm{h.c.}\,.
\end{eqnarray*}

Keeping only FCNC parts we obtain

\begin{eqnarray*}
O^{u}_{LL} &=&  \frac{(v+h)^2}{2}[g A_{\mu}^3  - g' B_{\mu}]\,\,\Big[ \bar{ t}_L \gamma^{\mu}c_L\Big] + \mathrm{h.c.} \,, \\
O_{LL}^h &=& \frac{(v+h)^2}{2}[g A_{\mu}^3  - g' B_{\mu}]\,\,\Big[\bar{t}_L \gamma^{\mu} c_L + \bar{b}_L \gamma^{\mu} s_L\Big] + \mathrm{h.c.} \,, \\
O_{RL}^w &=&g\frac{v+h}{\sqrt{2}}W_{\mu\nu}^3\,\,\Big[\bar{c}_L \sigma^{\mu\nu} t_R\Big] + \mathrm{h.c.} \,, \\
O_{RL}^b &=& g'\frac{v+h}{\sqrt{2}} B_{\mu\nu}\,\,\Big[\bar{c}_L \sigma^{\mu\nu} t_R\Big]+ \mathrm{h.c.} \,, \\
O_{LR}^w &=& g\frac{v+h}{\sqrt{2}}W_{\mu\nu}^3\,\,\Big[\bar{t}_L \sigma^{\mu\nu} c_R\Big] + \mathrm{h.c.} \,, \\
O_{LR}^b &=&g'\frac{v+h}{\sqrt{2}} B_{\mu\nu}\,\,\Big[\bar{t}_L \sigma^{\mu\nu} c_R\Big]+ \mathrm{h.c.} \,, \\
O_{RR}^u &=& \frac{(v+h)^2}{2}[g A_{\mu}^3  - g' B_{\mu}]\,\,\Big[\bar{t}_R\gamma^{\mu}c_R \Big] + \mathrm{h.c.} \,.
\end{eqnarray*}

The electroweak coupling constants are
\begin{equation*}
g_Z = \frac{2e}{\sin 2\theta_W}=\frac{g}{\cos\theta_W}\hspace{0.5cm} e = g \sin\theta_W \hspace{0.5cm} \frac{g'}{g} = \tan\theta_W\,,
\end{equation*}
where $\theta_W$ is the Weinberg angle.
Finally our coupling constants can be expressed as
\begin{eqnarray}
a_L^Z &=& \frac{1}{2}\Big[C^u_{LL}+ C_{LL}^h\Big]\,,\label{eq:fox1}\\
a_R^Z &=& \frac{C^u_{RR}}{2} \,, \label{eq:fox2}\\
b_{LR}^Z &=& \frac{C_{RL}^w \cos^2\theta_W - C_{RL}^b \sin^2\theta_W}{\sqrt{2} } \,,\label{eq:fox3}\\
b_{RL}^Z &=& \frac{C_{LR}^w \cos^2\theta_W -C_{LR}^b \sin^2\theta_W}{\sqrt{2}}\,,\label{eq:fox4}\\
b_{LR}^{\gamma} &=& \frac{C_{RL}^w + C_{RL}^b }{\sqrt{2}}\,,\label{eq:fox5}\\
b_{RL}^{\gamma} &=& \frac{C_{LR}^w  + C_{LR}^b }{\sqrt{2}}\,.\label{eq:fox6}
\end{eqnarray}
Scalar couplings $c_{LR}^h$ and $c_{RL}^h$ are not included here because the operators considered in ~\cite{Fox:2007in}
do not contain $tch$ vertices.

\bibliography{article}

\end{document}